\documentclass{aa}  
\defcitealias{Basu25}{B25}
\defcitealias{Khullar+20}{K20}
\usepackage{txfonts}

\usepackage{graphicx}
\usepackage{float}
\usepackage{newtxtext,newtxmath}
\usepackage[T1]{fontenc}
\usepackage{xcolor}
\usepackage{subfig}
\usepackage[normalem]{ulem}

\usepackage{hyperref}

\DeclareRobustCommand{\VAN}[3]{#2}
\let\VANthebibliography\thebibliography
\def\thebibliography{\DeclareRobustCommand{\VAN}[3]{##3}\VANthebibliography}

\usepackage{graphicx}	
\usepackage{amsmath}	
\usepackage{amssymb}	
\usepackage{siunitx}

\newcommand{\HeII}{He~{\small{II}} }
\newcommand{\HeIII}{He~{\small{III}} }

\usepackage{txfonts}
\usepackage{tikz,xcolor,hyperref}
\definecolor{lime}{HTML}{A6CE39}
\DeclareRobustCommand{\orcidicon}{%
	\begin{tikzpicture}
	\draw[lime, fill=lime] (0,0) 
	circle [radius=0.16] 
	node[white] {{\fontfamily{qag}\selectfont \tiny ID}};
	\draw[white, fill=white] (-0.0625,0.095) 
	circle [radius=0.007];
	\end{tikzpicture}
	\hspace{-2mm}
}

\foreach \x in {A, ..., Z}{%
	\expandafter\xdef\csname orcid\x\endcsname{\noexpand\href{https://orcid.org/\csname orcidauthor\x\endcsname}{\noexpand\orcidicon}}
}

\begin{document}

   \title{Detecting the signature of helium reionization through $^3\mathrm{He\small{II}}$ 3.46~cm line-intensity mapping}

   \subtitle{ }

    \author{Benedetta~Spina\inst{1,2}\fnmsep\thanks{b.spina@thphys.uni-heidelberg.de}\orcidA{},
          Cristiano~Porciani\inst{2,3,4,5}\orcidB{},
          Sarah~E.~I.~Bosman\inst{1,6}\orcidC{},
          Frederick~B.~Davies\inst{6}\orcidD{},
          Enrico~Garaldi\inst{7}\orcidE{},
          Ryan~P.~Keenan\inst{6}\orcidF{},
          \and
          Carlo~Schimd\inst{8,9}\orcidG{}
          }
   \authorrunning{Spina et al.}
   \institute{Institute for Theoretical Physics, Heidelberg University, Philosophenweg 12, D–69120, Heidelberg, Germany
        \and
        Universit\"at Bonn, Argelander-Institut f\"ur Astronomie, Auf dem H\"ugel 71, 53121 Bonn, Germany
            \and
            SISSA, International School for Advanced Studies, Via Bonomea 265, 34136 Trieste, TS, Italy
                \and
                Dipartimento di Fisica – Sezione di Astronomia, Università di Trieste, Via Tiepolo 11, 34131, Trieste, Italy
                    \and
                    IFPU, Institute for Fundamental Physics of the Universe, via Beirut 2, 34151 Trieste, Italy
                        \and
                        Max-Planck-Institut f\"{u}r Astronomie, K\"{o}nigstuhl 17, 69117 Heidelberg, Germany
                            \and
                            Kavli IPMU (WPI), UTIAS, The University of Tokyo, Kashiwa, Chiba 277-8583, Japan
                                \and
                                Aix Marseille Univ, CNRS, CNES, LAM, Marseille, France
                                    \and
                                    INAF – Astronomical Observatory of Trieste, via G.B. Tiepolo 11, I-34143 Trieste, Italy
                                 }

   \date{ }

\abstract
{Helium reionization is the most recent phase change of the intergalactic medium, yet its timing and main drivers remain uncertain. Among the probes to trace its unfolding, the 3.46\,cm hyperfine line of singly-ionized helium opens the study of helium reionization to upcoming radio surveys.}
{We aim to evaluate the detectability of the 3.46\,cm signal with radio surveys and the possible constraints it can place on helium reionization, in particular whether it can distinguish between early and late helium reionization scenarios. Moreover, we perform a comprehensive study of the advantages of single-dish vs.\ interferometric setup.}
{Using hydrodynamical simulations post-processed with radiative transfer, we construct mock data cubes for two models of helium reionization. We compute the power spectrum of the signal and forecast the signal-to-noise ratio for SKA-1 MID, DSA-2000, and a PUMA-like survey, in both observational setups.}
{The two scenarios produce distinct power spectra, but the faintness of the signal, largely caused by weak coupling between the spin temperature and the kinetic temperature in low-density regions of the IGM, combined with high instrumental noise, makes detection very difficult within realistic integration times for current surveys. A PUMA-like survey operating in single-dish mode could, however, detect the 3.46\,cm signal with an integrated signal-to-noise ratio of a few in $\lesssim 1000\,\mathrm{h}$ in both scenarios.}
{Distinguishing helium reionization scenarios with 3.46\,cm line-intensity mapping therefore remains challenging for current facilities. Our results, however, indicate that next-generation, high-sensitivity surveys with optimized observing strategies, especially when combined with complementary probes of the IGM, could begin to place meaningful constraints on the timing and morphology of helium reionization.}


   \maketitle
%

\section{Introduction}

The main components of the intergalactic medium (IGM) underwent two global changes in the first few billions years of the cosmic history, the ionization of neutral hydrogen \citep[H~{\small{I}} $\to$ H~{\small{II}}, for a review see][]{BarkanaLoeb2005}, and of helium. The first ionization of helium (He~{\small{I}} $\to$ He~{\small{II}}) has a ionization potential ($E_P^\mathrm{HeI} = 24.6$ eV) comparable to the one needed to ionize neutral hydrogen ($E_P^\mathrm{HI}$ is $13.6$ eV), while the second ionization of helium \citep[He~{\small{II}} $\to$ He~{\small{III}}, for a review see][]{Furlanetto+08,BaglaLoeb09,McQuinnSwitzer09} is characterized by a larger ionization potential ($E_P^\mathrm{HeII} = 54.4$ eV). 
While photons emitted by star-forming galaxies at high redshift ionize both neutral hydrogen and neutral helium, they do not significantly contribute to the second ionization of helium (hereafter referred to as just helium reionization), which requires photons with higher energies (as those emitted by active galactic nuclei, AGNs). For this reason, helium reionization is closely related to the abundance and properties of the quasar (QSOs) population in the early Universe and it can be use as a probe of QSOs activity and galaxy formation.

The majority of current studies agree on a Universe fully reionized at redshift $z \sim 3.2$. This picture is supported by evidences from the evolution of the IGM temperature from the H~{\small{I}} Ly$\alpha$ absorption features \citep[][]{ Theuns+02,Becker+11,Walther+19,Garzilli+20,Gaikwad+21}, from observations of the quasar luminosity function \citep[QLF,][]{Richards+06,Jiang+16}
and from the helium Ly$\alpha$ forest \citep{Jakobsen+94,Shull+10, Syphers+11,Worseck+16,Worseck+19}.

Measurements of the quasar luminosity function in large spectroscopic surveys \citep[e.g.][]{Croom+04, Richards+06,Jiang+16} combined with assumptions on their spectral energy distribution in the ultraviolet are in reasonable agreement with the picture described above:
the estimated ionizing emissivity of AGNs rapidly decreases with redshift at $z>3$. However, recent observational campaigns in different wavebands provided some evidence for a much larger population of ultra-bright quasars \citep{Grazian+22} and faint AGNs \citep[][but see however \citet{Ricci+17, Parsa+18}]{Giallongo+15} which may be able to trigger an earlier helium reionization and even contribute significantly to hydrogen reionization \citep{MadauHaart15,Chardin+17,Fontanot+23,Madau24}. Moreover, the James Webb Space Telescope provided strong evidence for a large population of obscured AGNs \citep[e.g.][]{Harikane+23, Yang+23}.
Depending on the opening angle of the dusty torus which causes the obscuration, hard-UV photons propagating along unobscured lines of sight could contribute to the reionization of both hydrogen and helium \citep[see in particular][]{Basu+24}.

\begin{figure}[ht!]
    \centering
    \includegraphics[width=\columnwidth]{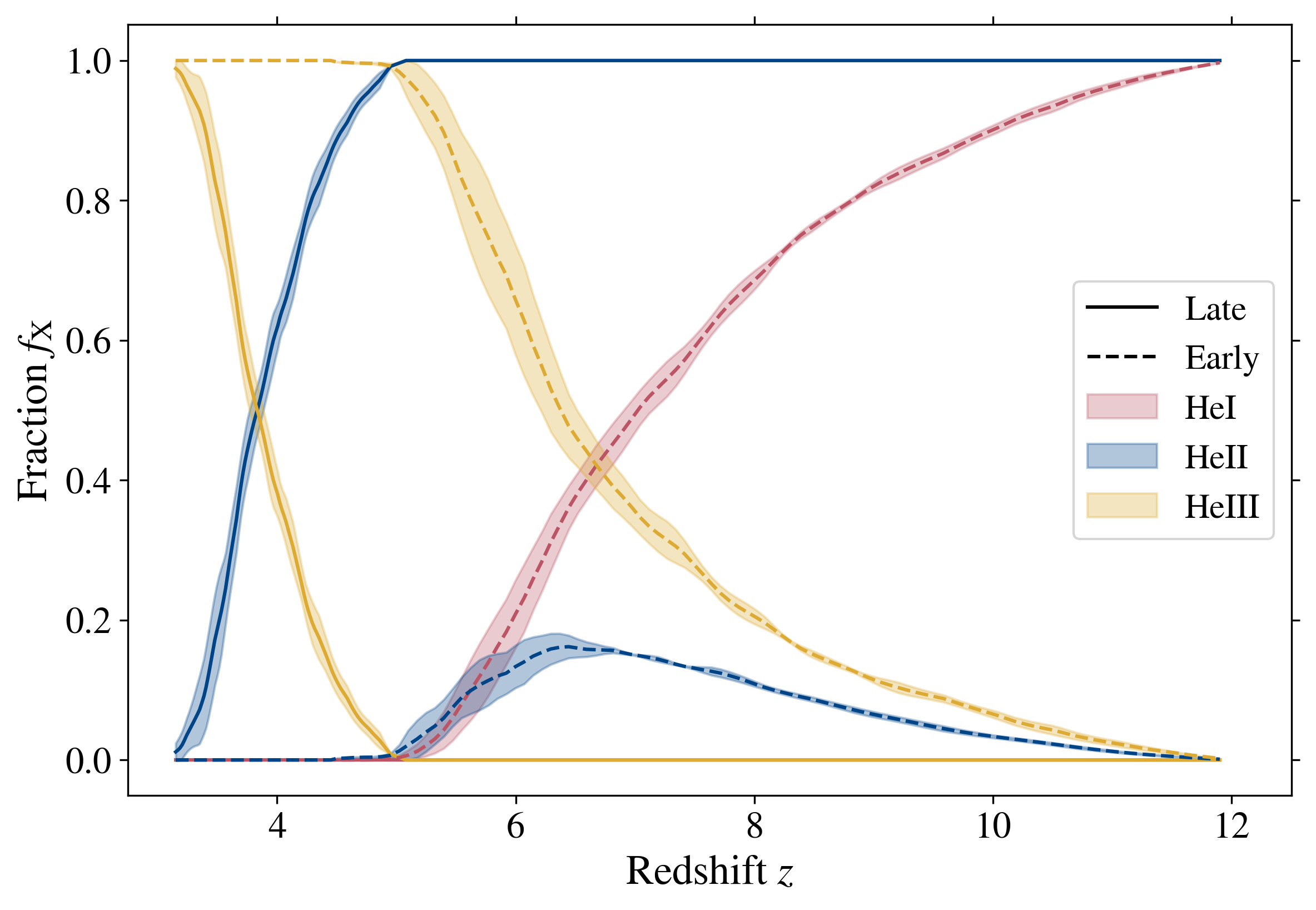}
    \caption{The He~{\small{I}} (red), He~{\small{II}} (blue), He~{\small{III}} (yellow) fractions for Late reionization (continuous curves) and Early reionization (dashed curves) as a function of redshift. It is noteworthy that $f_\mathrm{HeI} = 0$ consistently for the Late model, whereas in the Early model, the first and second ionization of helium occur simultaneously.}
    \label{fig:fig1}
\end{figure}

On small scales, theoretical studies on helium reionization from ionized zones near accreting black holes \citep[][]{Vasiliev+19,Worseck+21} or with fast radio bursts \citep[][]{Linder20} have been developed. 

We can investigate helium reionization through, for example, the hyperfine transition at $\lambda_\mathrm{HeII} = 3.46$~cm in singly ionized $^3$He atoms, analogous to the $\lambda_\mathrm{HI} = 21$~cm line for neutral hydrogen atoms \citep{BaglaLoeb09,McQuinnSwitzer09}. 

Helium is the second most abundant element in the Universe (with a primordial mass fraction of $Y = 0.24$), almost entirely in the form of $^4$He. The lighter isotope $^3$He is present only at a level of $\sim 10^{-5}$ relative to $^4$He, with a spontaneous decay rate for He~{\small{II}} $\sim 680$ times larger than the one for H~{\small{I}}.
As a result, the 3.46~cm emission line is more pronounced and spatially diffuse compared to, for instance, the [CII] or CO lines, and fainter than the H~{\small{I}} line. With a rest-frame frequency of $\nu_\mathrm{HeII} = 8.67$ GHz, it is less susceptible to contamination from synchrotron radiation and the terrestrial ionosphere than the 21~cm line \citep[although foregrounds still challenge the detection of a global signal, see e.g.][]{Takeuchi+14}.

At low redshift, detection of the 3.46~cm signal from local He~{\small{II}} regions and planetary nebulae were performed with radio telescopes such as the GBT \citep[][]{Rood+79,Rood+84,BalserBania18} and the VLA \citep[][]{Balser+06}. 

\citet{TrottWayth24} recently presented the first constraints on the cross-correlation power spectrum of He~{\small{II}} at $2.9 \leq z \leq 4.1$, from the Australia Telescope Compact Array\footnote{https://www.narrabri.atnf.csiro.au/}, as an upper limit on the brightness temperature fluctuations of the 8.67 GHz hyperfine transition, showing the potential of this technique. From a theoretical perspective, a similar approach has been proposed in \citet{Khullar+20} (hereafter \citetalias{Khullar+20}) forecasting for SKA-1 MID in interferometric setup and focusing on temperature fluctuations around QSOs at high redshift (with an early end of helium reionization), and more recently in \citet{Basu25} (hereafter \citetalias{Basu25}) where the 3.46cm forest \citep[analogue to the 21cm forest, e.g.][]{Furlanetto06} has been investigated as well.

The goal of several upcoming and ongoing radio surveys is to observe hydrogen and helium reionization and scrutinize their main drivers. Among these, the Hydrogen Intensity and Real-time Analysis eXperiment \citep[HIRAX,][]{hirax}, \citep[CHIME,][]{Bandura+14}, the LOw Frequency ARray \citep[LOFAR,][]{Haarlem+13}, the Square Kilometre ArrayPhase 1 MID \citep[SKA-1 MID,][]{SKAredbook}, the Deep Synoptic Array 2000 \citep[DSA-2000,][]{Hallinan+19} and the Packed Ultra-wideband Mapping Array \citep[PUMA,][]{Slosar+19}.
In the study of hydrogen reionization, it is well established \citep{Battye+13, Bull+15} that radio surveys operating in single-dish mode are optimal for constraining cosmological scales at low redshift (i.e., in the post-reionization era), whereas interferometers are more sensitive to small-scale fluctuations and perform better at higher redshift. However, the optimal configuration for studying helium reionization remains uncertain. 

The aim of this work is twofold: (i) to determine which observational setup is best suited for probing helium reionization through fluctuations in the 3.46~cm helium line, and (ii) to assess the detectability of this signal using SKA-1 MID, DSA-2000, and PUMA-like surveys, with the goal of constraining the nature of the sources driving helium reionization. We consider here: i) a standard late-reionization scenario, and ii) an early-reionization scenario, where a population of QSOs at high redshift is responsible for both hydrogen and helium reionization \citep{Giallongo+15}. We employ hydrodynamical simulations post-processed to solve the radiative-transfer equations and simulate radio-like data cubes. We account for instrumental effects, such as angular resolution and thermal noise. 

We focus on the 3D power spectrum rather than on the global signal, as fluctuations
constitute a more robust observable in the presence of bright smooth-spectrum
foregrounds, which still dominate over the He\,\textsc{ii} line, though less severely
than for the H\,\textsc{i} 21\,cm signal.

The work is organized as follows. In section \ref{sec:datacube} we describe the hydrodynamical simulations and explain how we generate mock data cubes. We address the issue of evaluating the brightness temperature of the signal, and its spin temperature in particular, in Sections \ref{sec:brightnesstemperature} and \ref{sec:spin_temperature}. The definition of the power spectrum and its computation from simulations are presented in Section \ref{sec:powerspectrum}. We introduce the radio surveys considered in this work in Section \ref{sec:surveys}, along with the noise associated to the detector for single-dish (Sections \ref{sec:freq_band}-\ref{sec:thermal_noise_sd}) and interferometer (Section \ref{sec:thermal_noise_int}) configurations, carefully selecting the appropriate wavelengths for each setup (Section \ref{sec:kmodes}).  The definition of signal-to-noise ratio and the results are presented and discussed in Sections \ref{sec:signal-to-noise} and \ref{sec:results}.

\section{Mock 3.46 cm data cube} \label{sec:datacube}
In this section, we describe the simulated mock \HeII data cubes and the calculation of the He~{\small{II}} brightness temperature in the hydrodynamical simulations. 

\subsection{Simulations} \label{sec:simulations}

This study is based on two suites of hydrodynamical simulations of the IGM performed with the \textsc{Ramses} code \citep{Teyssier02} and post-processed with the \textsc{Radamesh} code \citep{CantalupoPorciani11}. The simulated volume consists of a periodic cubic box of comoving side $L = 100 \,h^{-1}$\,Mpc, discretized in post-processing into a regular Cartesian mesh with $(N_\mathrm{grid})^3 = 128^3$ elements, and assumes a flat $\Lambda$CDM cosmology.  

In the first simulation suite\footnote{The following cosmological parameters are adopted: matter density $\Omega_{\rm m} = 0.2726$, baryon density $\Omega_{\rm b} = 0.0456$, Hubble constant $H_0 = 100\,h\,{\rm km\,s^{-1}\,Mpc^{-1}}$ with $h = 0.704$, spectral index $n_{\rm s} = 0.963$, and linear rms fluctuation $\sigma_8 = 0.809$ within spheres of radius $8\,h^{-1}{\rm Mpc}$.} \citep[][hereafter \textit{Late}]{Compostella+14}, consisting of six realizations, helium reionization is driven by AGNs that turn on at redshift $z = 5$, when both hydrogen and helium are already singly ionized.  
The underlying hydrodynamical run evolves from $z = 120$ to $z = 4$, and the radiative--transfer calculation is then performed from $z = 5$ down to $z \simeq 2.5$ on this fixed density field, allowing only the ionization and temperature to evolve.  
A stochastic algorithm associates dark-matter haloes with AGNs in a way that reproduces the observed luminosity function of \citet{Glikman+11}.  
A uniform, time-varying UV background mimics the contribution from star-forming galaxies.  
As a result, helium reionization is completed by $z \sim 3$ (late reionization scenario).

The second simulations suite\footnote{With cosmological parameters: $\Omega_{\rm m} = 0.306$, $\Omega_{\rm b} = 0.0483$, $H_0 = 100\,h\,{\rm km\,s^{-1}\,Mpc^{-1}}$ with $h = 0.679$, $n_{\rm s} = 0.958$, and $\sigma_8 = 0.815$.} \citep[][hereafter Early]{Garaldi+19}, consisting of four realizations, incorporates the QSO luminosity function from \citet{Giallongo+15} with a large population of QSOs at high redshift and corresponds to an early-reionization model. The luminosity of the QSOs --- placed at the center of the simulated dark-matter haloes --- is calibrated using the \citet{Giallongo+15} QSO luminosity function at $z=4$ and extrapolated to higher redshift following \citet{Madau15}; no UV radiation background is included (ionizing photons emitted by stars are not considered). A lightbulb model is used to represent QSO activity, where sources randomly activate with a probability independent of the host halo's properties. In this case, QSOs provide enough power to ionize both H~{\small{I}}, He~{\small{I}} and He~{\small{II}}, resulting in the end of helium reionization much earlier than in standard models and almost coincident with the end of hydrogen reionization, at $z \approx 5$.

Figure \ref{fig:fig1} presents the evolution of the average fraction of He~{\small{I}} (red), He~{\small{II}} (blue), and He~{\small{III}} (yellow) for the Late model (continuous curves) and the Early model (dashed curves), the shaded regions indicating the standard deviation among the simulation runs. In the Late model, helium is already ionized once at the beginning of the simulation ($f_\mathrm{HeI} = 0$ for $3 \lesssim z \lesssim 5$ while $f_\mathrm{HeII} = 1$ at $z = 5$). In the Early model, however, the first and second ionization of helium occur simultaneously. The majority of singly-ionized helium is immediately ionized again, causing the He~{\small{II}} fraction to slowly increase up to $f_\mathrm{HeII} \lesssim 0.2$ at $z \sim 6.5$ before subsequently decreasing. By the end of both simulations, respectively at $z \sim 3$ and $z \sim 5$, helium is completely ionized.

\begin{figure}[ht!]
    \centering
    \subfloat[Late model\label{fig:Tspin_late}]{\includegraphics[width=\columnwidth]{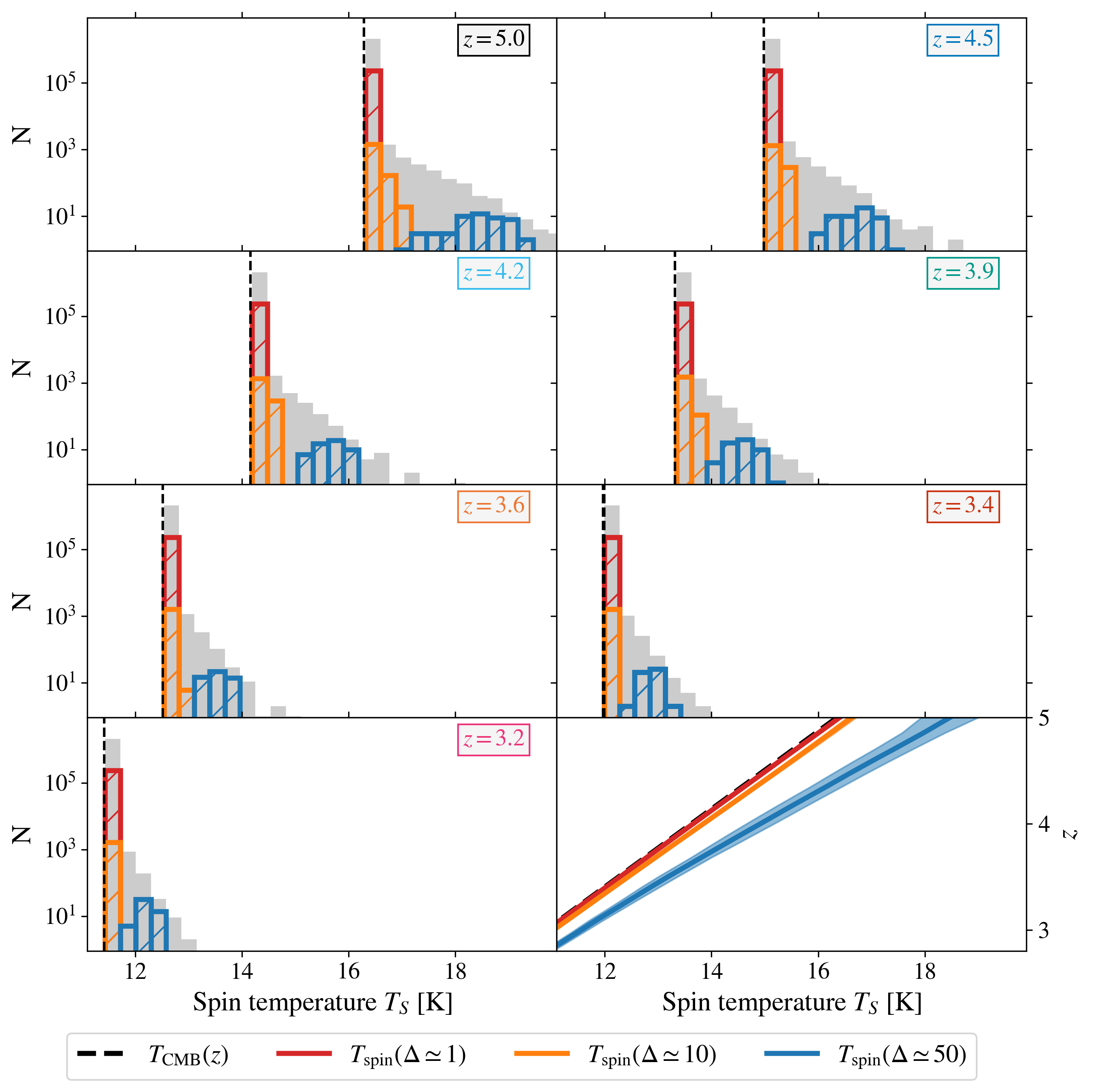}} \\
    \subfloat[Early model\label{fig:Tspin_early}]{\includegraphics[width=\columnwidth]{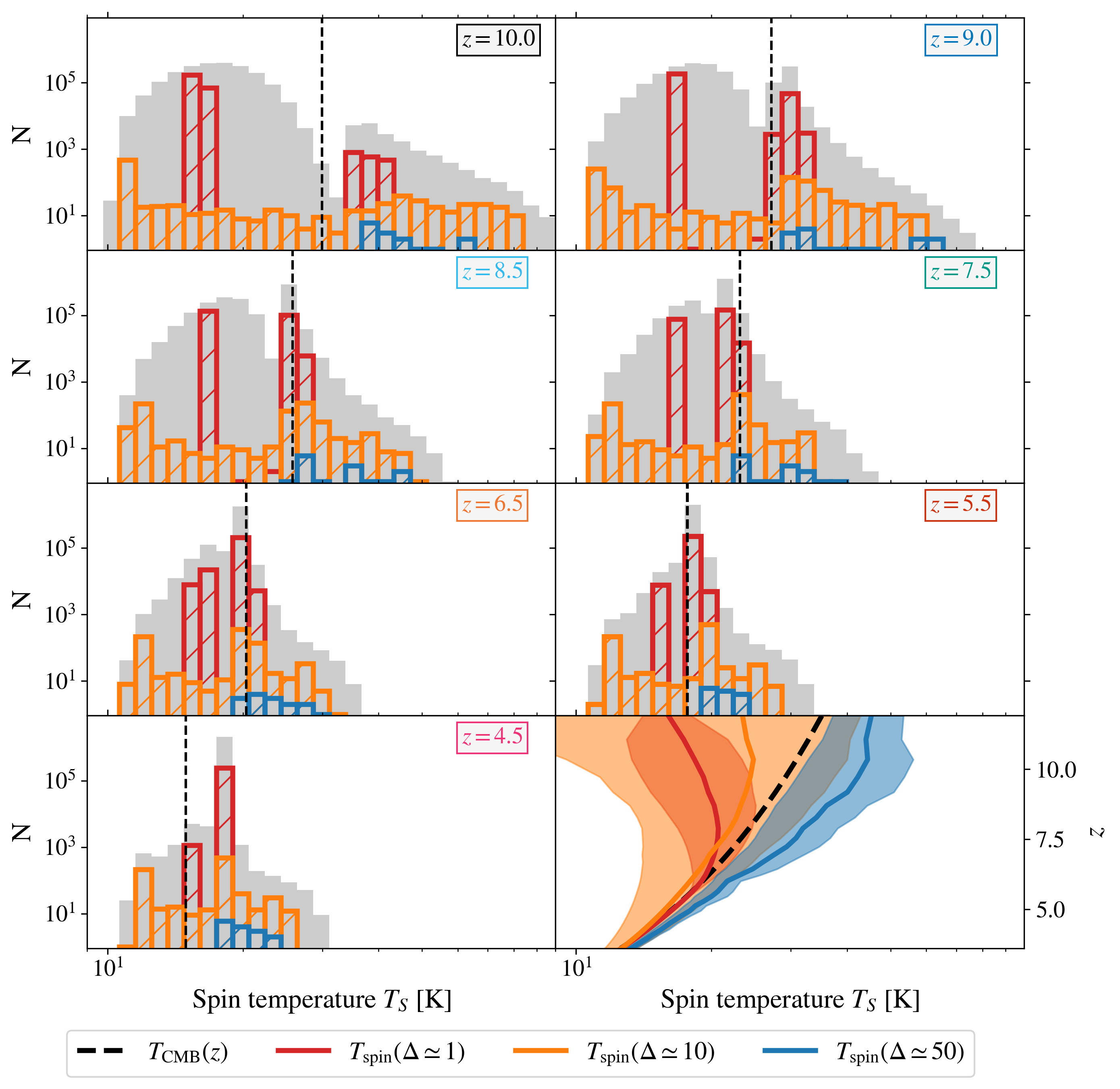}}
    \caption{Spin temperature distribution for both models at various redshifts. For each snapshot, the panels show the number of simulation voxels whose spin temperature falls within a given bin, $N$, as a function of the spin temperature.
    The spin temperature values corresponding to three overdensities, $\Delta_b = 1, 10, 50$, are highlighted in red, orange and blue. The bottom-right panels display the evolution of the mean spin temperature with redshift together with its standard deviation.}
    \label{fig:spin}
\end{figure}

\begin{figure*}[h!]
    \centering
    \centerline{\includegraphics[width=\textwidth]{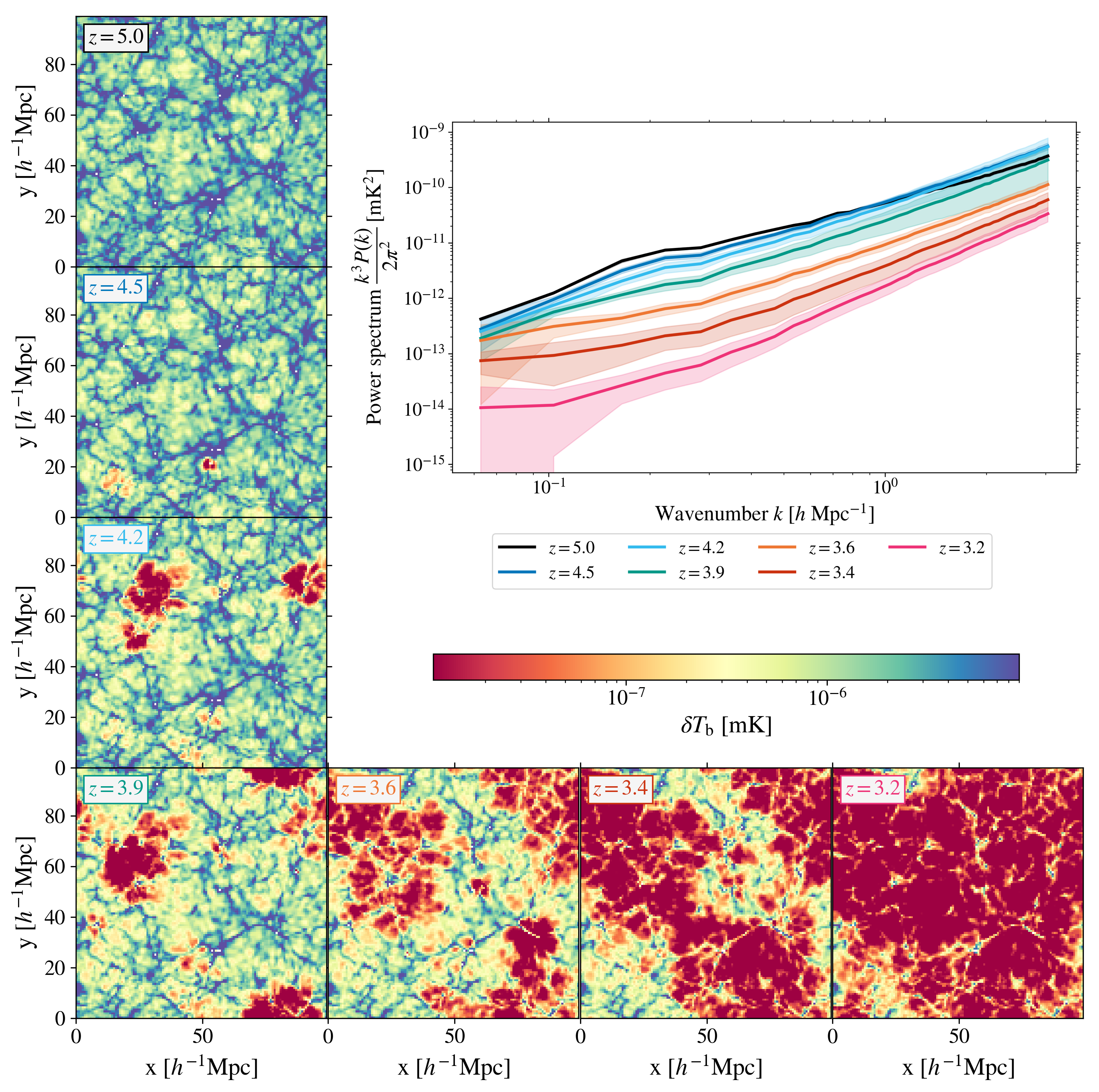}}
    \caption{Maps of the brightness temperature for the Late model at various redshifts, accompanied by the corresponding power spectra. The continuous curves represent the mean power spectra in the simulations, while the shaded regions show the $1\sigma$ scatter around the mean.}
    \label{fig:fig2}
\end{figure*}

\begin{figure*}[h!]
    \centering
    \centerline{\includegraphics[width=\textwidth]{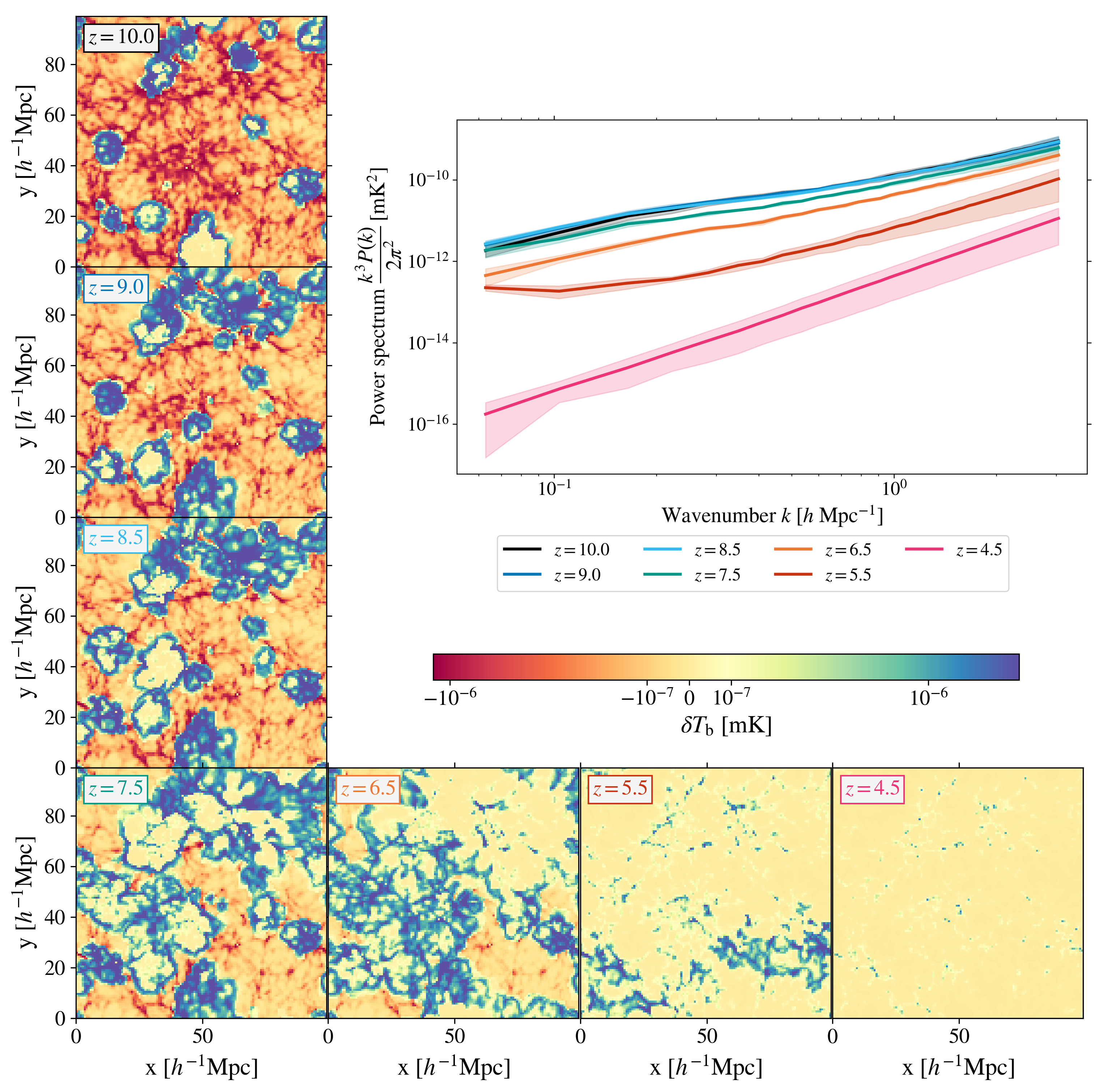}}
    \caption{As for Figure \ref{fig:fig2} (to enhance the morphology of the signal, a different color scale for the brightness temperature is used; similarly for the power spectrum measurement), assuming the Early reionization scenario. }
    \label{fig:fig3}
\end{figure*}

\subsection{Brightness temperature}\label{sec:brightnesstemperature}

The differential brightness temperature $\delta T_b$ of the 3.46~cm hyperfine transition of He~{\small{II}} with respect to the CMB is given by
\begin{equation}
    \delta T_b(z,\mathbf{r}) = \dfrac{T_S(z,\mathbf{r}) - T_\mathrm{CMB}(z)}{1+z}\left[1-e^{-\tau(z,\mathbf{r})}\right],
\end{equation}
where $T_S (z,\mathbf{r})$ is the spin temperature associated to the spin-flip of the electron in the He~{\small{II}} atoms, at redshift $z$ and position $\mathbf{r}$, $T_\mathrm{CMB} (z)$ the CMB temperature and $\tau (z,\mathbf{r})$ the helium optical depth. In the optically thin regime $\tau \ll 1$, this corresponds to \citep{Furlanetto+06}:
\begin{align}
    \nonumber \delta T_b(z,\mathbf{r}) & \simeq \dfrac{T_S(z,\mathbf{r}) - T_\mathrm{CMB}(z)}{1+z}\tau(z,\mathbf{r}) \simeq \\
    & \simeq 0.58\, f_\mathrm{HeII}(z,\mathbf{r})\Delta_b (1+z)^2 \frac{H_0}{H(z)} \label{eqn:dtb} \\ 
    \nonumber & \hspace{2cm} \times \left[ 1 - \dfrac{T_\mathrm{CMB}(z)}{T_S(z,\mathbf{r})}\right] \left[ \dfrac{H(z)/(1+z)}{\mathrm{d}v_\parallel / \mathrm{d}r_\parallel} \right] \,\mu\mathrm{K}, 
\end{align} 
where $f_\mathrm{HeII}$ is the local fraction of He~{\small{II}}, $\Delta_b (=\rho_b/\rho_c)$ the matter overdensity and $\mathrm{d}v_\parallel / \mathrm{d}r_\parallel$ as the velocity gradient along the line-of-sight, including both the Hubble expansion and the peculiar velocity. In this work we assume that $ \frac{H(z)/(1+z)}{\mathrm{d}v_\parallel / \mathrm{d}r_\parallel} \sim 1$. 

\subsection{Spin temperature} \label{sec:spin_temperature}
Three processes contribute to the spin temperature: i) the coupling to the CMB temperature, ii) the collisional (de-)excitation, iii) the absorption and re-emission of Lyman-series photons \citep{Field58, Takeuchi+14}. As a first approximation, the spin temperature is given by the equilibrium of these processes, 
\begin{equation}
T_S = \dfrac{T_\mathrm{CMB} + y_{\alpha}T_{\alpha} +  y_{c}T_{K}}{1 + y_{\alpha} + y_c},
\label{eqn:spin_temperature}
\end{equation}
where $T_{\alpha}$ the temperature of the Ly-$\alpha$ radiation, $T_K$ is the gas kinetic temperature, closely coupled with $T_{\alpha}$ by recoil during repeated scattering and with the gas temperature (i.e. $T_g \simeq T_K \simeq T_{\alpha}$), $y_c$ is the coupling coefficient related to atomic collisions and $y_{\alpha}$ is the coupling coefficient due to the scattering of Ly-$\alpha$ photons.

Both coupling coefficients are proportional to the rest-frequency of the emission line and to the inverse of the spontaneous decay rate, boosting the signal with respect to the 21 cm line from neutral hydrogen. However, both the collisional de-excitation rate and the de-excitation driven by
He\,\textsc{ii} Lyman-$\alpha$ scattering are extremely small for the ${}^3$He$^{+}$ hyperfine transition. The dominant limitation is the very low abundance of ${}^3$He ($n_{^3\mathrm{He}}/n_{\mathrm{H}}\sim10^{-5}$), which strongly suppresses both collisional and radiative coupling irrespective of the detailed cross-sections. In addition, the relevant spin–exchange cross-sections and the background flux at the He\,\textsc{ii} Lyman-$\alpha$ wavelength are themselves small, further reducing the efficiency of these processes.   As a result, efficient coupling occurs only in the densest and coolest regions of the IGM, while the diffuse IGM remains only weakly coupled.

Contrary to \citetalias{Khullar+20} and \citetalias{Basu25}, where the spin temperature is approximated to the gas temperature,\footnote{\citetalias{Khullar+20} focused mainly on the temperature fluctuations around QSOs (i.e. in overdense regions, where $T_S \gg T_\mathrm{CMB}$).} we compute the spin temperature following equation \ref{eqn:spin_temperature} \citep[][]{Takeuchi+14}, i.e.~accounting for the coupling with both atomic collisions and scattering of Ly-$\alpha$ photons. We assume the model of UV/X-ray background from \citet{HaardtMadau12} for the Ly-$\alpha$ flux as a function of redshift. 

Figures \ref{fig:Tspin_late} and \ref{fig:Tspin_early} show the spin temperature for different redshifts for each model. It is worth noting that $T_S \geq T_\mathrm{CMB}$ at all redshifts in the Late model, but not for the Early one, where a mixture of neutral, single-ionized and double-ionized helium is always present. Moreover, the spin temperature distribution considering three overdensities, $\Delta_b = 1, 10, 50$, in red, orange and blue, in shown. As expected, in both models the spin temperature is higher for larger overdensities, making $\delta T_b$ an optimal tracer of the underlying matter distribution.

Being the spin temperature highly coupled with the CMB, except in the high-density regions discussed above, it is not possible to generally neglect the CMB temperature with respect to the spin temperature, resulting in $\left(1 - T_\mathrm{CMB}/T_S\right) \not\approx 1 $ in Eq.~\ref{eqn:dtb}.

Unresolved small--scale density structure below the spatial resolution of our simulations could, in principle, enhance the 3.46\,cm signal. In the fully coupled regime the differential brightness temperature scales approximately linearly with overdensity, $\delta T_b \propto \Delta_b$, so sub–grid clumping does not alter the mean signal. A possible boost could arise if the factor $(1 - T_{\rm CMB}/T_S)$ exhibited a strong density dependence, since in that case $\langle \delta T_b(\Delta_b) \rangle \neq \delta T_b(\langle \Delta_b \rangle)$, analogous to the clumping corrections for processes that scale as $\Delta_b^2$. However, in the redshift range relevant for this work (prior to the completion of helium reionization), the dependence of $T_S$ on overdensity is only mildly super–linear \citep{Takeuchi+14}, and becomes significant only at lower redshifts where the He\,\textsc{ii} fraction is already negligible. Therefore, sub–Mpc clumping is expected to induce at most a very small enhancement of the He\,\textsc{ii} hyperfine signal.

\section{Power spectrum} \label{sec:powerspectrum}
Denoting by $\delta T_b(\mathbf{q})$ the Fourier transform of the differential brightness temperature, where $\mathbf{q}$ are wavevectors with Cartesian components integer multiples of the fundamental frequency $k_\mathrm{F} = 2\pi/L$ (for this work, $k_\mathrm{F} = 0.0628 \,h\, \mathrm{Mpc}^{-1}$), the estimator of its power spectrum is
\begin{equation}
    \hat{P} (k) = \dfrac{1}{N_q} \sum_{\mathbf{q} \in k} |\delta T_b(\mathbf{q})|^2,
\end{equation}
where $N_q$ the number of $\mathbf{q}$ vectors in the $k$-bin of width $\delta k$ and the sum is run over the $\mathbf{q}$ vectors such as $k-\delta k/2 \leq |\mathbf{q}| < k+\delta k/2$.

The mean power spectrum of the He~{\small II} signal, along with the standard deviation over the set of simulations, is shown for the Late model in Figure~\ref{fig:fig2} and for the Early model in Figure~\ref{fig:fig3}, together with the corresponding brightness temperature field from a representative simulation run. A clear distinction emerges when comparing the signal strengths of the two models at the same redshift. This difference arises from the distinct reionization timelines adopted in the Late and Early models. These results underscore how the timing of helium reionization strongly influences the amplitude and evolution of the He~{\small II} signal as reflected in the power spectrum.

As reionization progresses, the bubbles expand, merge, and fill the intergalactic medium. This expansion creates distinct patterns in the power spectrum on large scales. The growing bubbles generate regions with enhanced He~{\small{II}} fraction surrounded by regions with lower He~{\small{II}} fraction, introducing fluctuations in the signal and changing the topology of the He~{\small{II}} field, transitioning from isolated and small bubbles to interconnected structures. In the Late model, where helium reionization occurs later and helium has already been ionized once, the He~{\small{III}} bubbles dominate and contain negligible He~{\small{II}} fraction inside them. The signal is produced outside the bubbles. In the Early model, helium is neutral at the beginning of the simulation, resulting in the formation of bubbles containing both He~{\small{II}} and He~{\small{III}}. Inside the bubbles, He~{\small{III}} dominates, while the He~{\small{II}} fraction is concentrated at the boundaries. Thus, the He~{\small{II}} signal mainly arises from the edges of such bubbles. 

The brightness temperature of the 3.46~cm signal from equation \eqref{eqn:dtb} is determined by both the \HeII fraction and the underlying matter density distribution. As such, we expected the power spectrum of the signal to reflect contributions from both components (Figure~\ref{fig:fig4} shows a comparison between the He~{\small{II}} power spectrum and the rescaled matter power spectrum at the same redshift). However, because the spin temperature plays a dominant role in shaping the brightness temperature (effectively suppressing the signal in low-density regions) the resulting 3.46~cm power spectrum more closely resembles the shape of the matter power spectrum than initially anticipated. While the amplitude differs due to fixed scaling constants, the imprint of ionized bubble structures is less apparent than expected.

Some evidence of deviation from this behavior emerges in the Late model when comparing the power spectra at redshifts $z=5$ and $z=4.5$. At $z=5$, the 3.46~cm power spectrum closely follows that of the matter distribution. At $z=4.5$, however, we observe a modest enhancement at high $k$-modes (corresponding to small scales), likely reflecting the contribution from \HeII ionized bubbles. At low $k$-modes (large scales), the power spectrum remains consistent with the matter distribution, suggesting that the bubble signal has only a limited impact on the overall structure at those scales.

Our forecasts yield power-spectrum amplitudes that are roughly two orders of magnitude lower than those reported by \citetalias{Basu25}, who obtained $\Delta(k{=}0.1\,h\,\mathrm{Mpc}^{-1}) \simeq 10^{-1}\,\mu\mathrm{K}$ for the
${}^3$He$^{+}$ hyperfine transition at $z\simeq4$.
At the same scale and redshift, we find $\Delta(k{=}0.1\,h\,\mathrm{Mpc}^{-1}) \simeq 10^{-3}\,\mu\mathrm{K}$
(corresponding to $\Delta^2\simeq10^{-6}\,\mu\mathrm{K}^2$).
This discrepancy originates from our self--consistent treatment of the spin--temperature coupling in the He\,\textsc{ii}
hyperfine signal: unlike \citetalias{Basu25}, who effectively assume $T_{\mathrm{S}}\!\gg\!T_{\mathrm{CMB}}$, we compute
$T_{\mathrm{S}}$ from collisional and Ly$\alpha$ coupling, which keeps it only slightly above the CMB temperature over most of the IGM
(see Sect.~\ref{sec:spin_temperature}). As a result, the factor $\left(1-T_{\mathrm{CMB}}/T_{\mathrm{S}}\right)$ remains well below unity (Fig.~\ref{fig:spin}), suppressing the amplitude of
$\delta T_{\mathrm{b}}$ and hence the power spectrum by roughly two orders of magnitude.

A qualitative interpretation of the \HeII power spectrum in terms of clustering and shot–noise regimes is provided in Section~\ref{sec:clustering}.


\section{Forecast for upcoming surveys}\label{sec:surveys}

We forecast helium reionization detectability for three radio surveys, operating in single-dish or interferometric configurations. The main technical features of the surveys are listed in Table \ref{tab:radiosurveys}. 

The first one we consider is the Square Kilometre Array Phase 1 MID (SKA-1 MID), a precursor to the full Square Kilometre Array (SKA) project. It will consist of an array of mid-frequency radio telescopes and it will have a large collecting area and high sensitivity, making it well-suited for studying the EoR and other cosmological open questions \citep{SKAredbook}.

The second radio survey we take into account is the Deep Synoptic Array 2000 (DSA-2000), a proposed radio survey telescope with $2000\times 5$m dishes. It aims to image the entire sky repeatedly over sixteen epochs, detecting over one billion radio sources and operating as a dedicated survey telescope \citep{Hallinan+19}.

Finally, the Packed Ultra-wideband Mapping Array (PUMA) is a proposed radio telescope designed for 21 cm intensity mapping in the post-reionization era, as well as other science goals like Fast Radio Bursts (FRBs), pulsar monitoring, and multi-messenger observations of transients. In the first phase, it consists of 5000x6m antennas \citep{Slosar+19}.

The redshift range target by these instrument, when looking at the 3.46~cm hyperfine transition, overlap with helium reionization completely for SKA-1 MID and DSA-2000 but only partially for PUMA. 
However, PUMA is the most advanced of the instruments considered, owing to its very large number of antennas and dense configuration, which make it particularly interesting for studies of helium reionization.
For this reason, as a proof of concept, we consider here a PUMA-like survey, i.e. a survey technically identical to PUMA but with a frequency bandwidth compatible with helium reionization. 

\begin{table}
\begin{center}
\begin{tabular}{c | c c c} 
 & SKA-1 MID & DSA-2000 & PUMA \\  \hline\hline
$D$ [m] & 15 & 5 & 6 \\ \hline
$N_\mathrm{dish}$ & 197 & 2000 & 5000 \\ \hline
$f_\mathrm{detec}$ [GHz] & 0.35-15.3 & 0.7 - 2.0 & 0.2 - 1.1 \\ \hline
$T_\mathrm{sys} (\nu)$ [K] & $\sim 25-80$ & $\sim 25$ & $\sim 25$  \\ \hline
$\delta \nu$ [kHz] & 13.4 & 162.5 & - \\ \hline
$B$ [km] & 1 & 4 & 0.6 \\
\end{tabular}
\caption{Technical parameters of the radio surveys considered in this work for the purposes of He\,\textsc{ii} detection. Here, $D$ is the diameter of the telescope antenna, $N_{\rm dish}$ the number of antennas, $f_{\rm detec}$ the detectable frequency range, $T_{\rm sys}$ the system temperature, $\delta\nu$ the frequency resolution, and $B$ the core baseline.}
\label{tab:radiosurveys}
\end{center}
\end{table}

\begin{figure}
    \centering
    \includegraphics[width=\columnwidth]{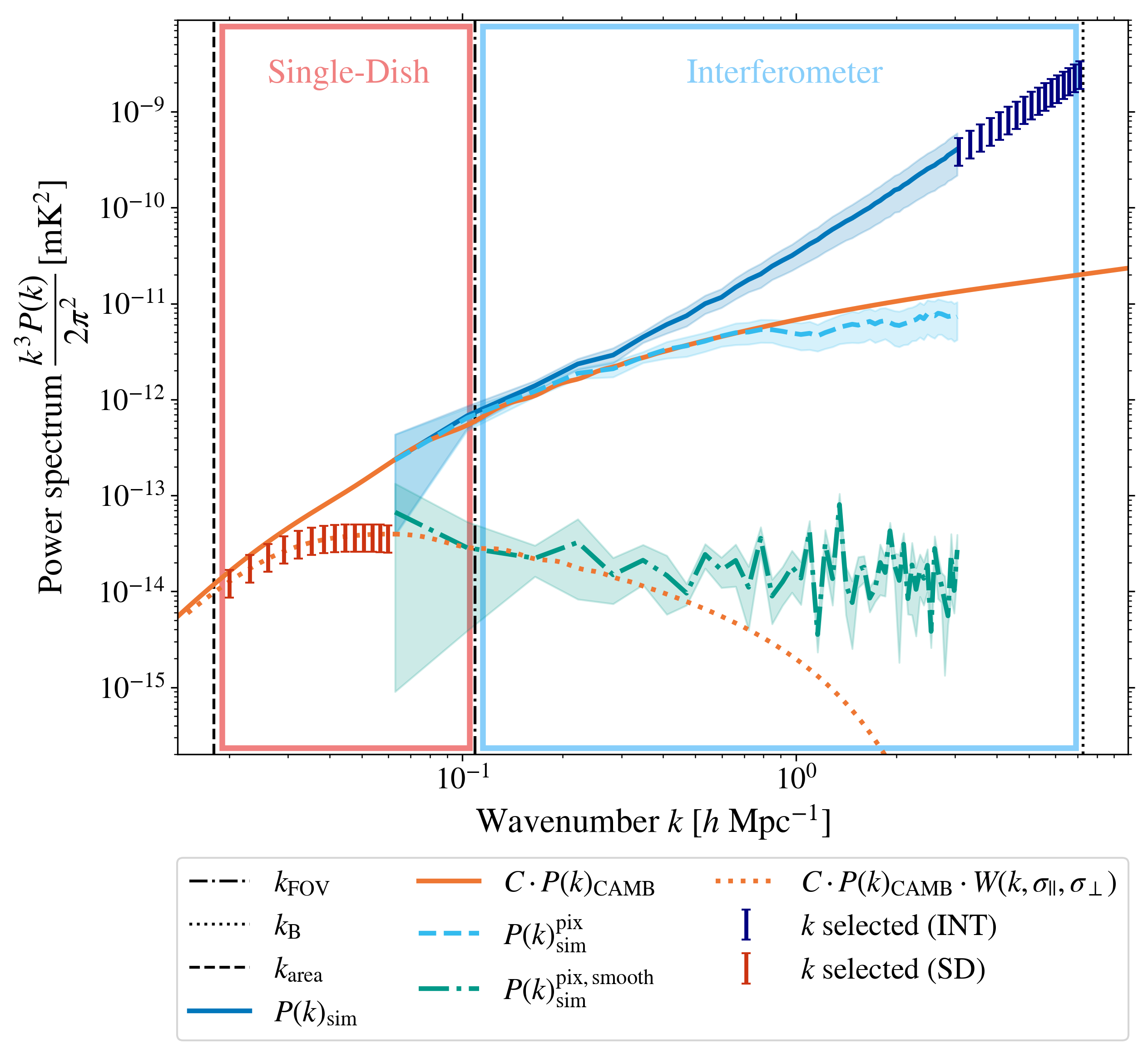}
    \caption{Analysis of the power spectrum of the 3.46 cm line signal at redshift z=4 for the Late reionization scenario. Shown are the measured power spectrum from the simulations (blue solid line), the spectrum after accounting for the finite pixel size (light-blue dashed line), and the spectrum after Gaussian smoothing due to the angular resolution (green dash-dotted line). The rescaled matter power spectrum is displayed both in its raw form (orange solid line) and after smoothing (orange dotted line). Shaded regions indicate the $1\sigma$ scatter across the different simulation runs. The $k$-range accessible to SKA1-MID in single-dish (interferometic) mode is highlighted in red (blue). Vertical reference scales denote the wavenumbers corresponding to the telescope's maximum angular resolution ($k_\mathrm{B}$, black-dotted line), its field of view ($k_\mathrm{FOV}$, black dash-dotted line), and its pointing area ($k_\mathrm{area}$, black dashed line). The Gaussian smoothing applied corresponds to the angular resolution of SKA1-MID at $z=4$.
    }
    \label{fig:fig4}
\end{figure}

We take into consideration the various systematic effects that arise during observations of the 3.46~cm hyperfine transition brightness temperature. As we are evaluating the surveys' performances in both single-dish (SD) and interferometric (INT) configurations, we describe here the systematic effects in both cases. Regarding the single-dish configuration, we account for various factors including the frequency bandwidth selection (Section \ref{sec:freq_band}), the angular resolution of the telescope (Section \ref{sec:ang_rel}), the foreground removal procedure (Section \ref{sec:foregrounds}), and the thermal noise associated with the instrument (Section \ref{sec:thermal_noise_sd}). The impact of these effects on the measured power spectrum are shown in Figure \ref{fig:fig4}, where the measured power spectrum (in the Late scenario at redshift $z = 4$) is reported for reference as blue continuous line. The thermal noise of the detector in an interferometric configuration differs from the one considered in a single-dish setup and it is described in Section \ref{sec:thermal_noise_int}.


\subsection{Frequency bandwidth and pixel size} \label{sec:freq_band}

When performing observations, the total receiver bandwidth is partitioned into frequency channels. 
Narrower channels offer better spectral resolution but lower sensitivity, while wider channels sacrifice spectral resolution for improved signal-to-noise ratio and sensitivity to large-scale structures. Selecting the optimal channel bandwidth involves balancing these factors based on the scientific goals and instrumental constraints of the tomography. The relation between redshift and comoving distance allows us to determine the channel bandwidth $\Delta \nu$ at a specific redshift $z$ corresponding to a given radial separation in comoving distance $\Delta \chi$, as
\begin{equation} \label{eqn:deltachi}
\Delta \chi
= \int_{z_{-}}^{z_{+}} c \dfrac{\mathrm{d} z}{H(z)}, \qquad z_{\mp}= \dfrac{1}{\displaystyle{\frac{1}{1+z}\pm\frac{\Delta \nu}{2\nu_{\mathrm{rest}}}}}-1\;.
\end{equation}
Accounting for the Cartesian grid size of the simulations, we set $\Delta \chi = L/N_\mathrm{grid} = 0.78\, h^{-1} \mathrm{Mpc}\, (\simeq 0.5 \mathrm{\,MHz\,at\,} z=4)$, i.e. the smallest comoving distance accessible in the direction parallel to the line-of-sight within the simulation (verified to be larger than the frequency resolution of each instrument).


\subsection{Angular resolution} \label{sec:ang_rel}
To account for the finite angular resolution of the instrument, when operating in a single-dish setup, we incorporate a two-dimensional Gaussian filter in the observational plane, which we convolve with the simulated data cube in Fourier space. The full width at half maximum (FWHM) of the Gaussian beam is given by\footnote{Ideal illumination gives a coefficient of $1.02$ \citep[e.g.][]{Hall05,Chen+20},
while tapering and sidelobes can raise it to $\sim 1.4$.
We adopt $1.2$ as a standard effective value \citep[][]{Spina+21}.}
\begin{equation}
    \theta_\mathrm{FWHM} = 1.2\, \frac{\lambda_\mathrm{HeII}}{D} (1+z),
\end{equation}
where $D$ is the diameter of the telescope antenna, reported in Table \ref{tab:radiosurveys} for each telescope. This corresponds to an isotropic Gaussian smoothing with standard deviation
\begin{equation}
\Sigma= \dfrac{\theta_{\rm FWHM}\, (1+z)\,d_{\mathrm{a}}}{2\sqrt{2\ln 2}}\;,
\label{eqn:Sigma}
\end{equation}
where $d_{\mathrm{a}}$ is the angular-diameter distance to redshift $z$ in the background (in a flat universe, $(1+z)\,d_{\mathrm{a}}=\chi$). It follows that $\theta_\mathrm{FWHM} \approx 0.79$ deg and $\Sigma \approx 29.3 \,h^{-1}\mathrm{Mpc}$ at $z=4$ for e.g. SKA-1 MID.

In the direction perpendicular to the line-of-sight, the solid angle covered by each pixel is $\Omega_\mathrm{pix} \simeq \theta_\mathrm{pix}^2$.
To ensure proper sampling of the map and satisfy the Nyquist-Shannon theorem, the angular size of each pixel
should be smaller than $\theta_{\mathrm{FWHM}}$. In practical terms, $\theta_\mathrm{pix}$ is typically chosen to be between $\theta_{\mathrm{FWHM}}/7$ and $\theta_{\mathrm{FWHM}}/3$ \citep{marr2015fundamentals}. This range ensures that the map is adequately sampled, capturing the necessary level of detail without introducing significant aliasing effects.

The effects on the measured power spectrum when considering the proper pixel size and the telescope beam are shown respectively as the light blue dashed line and the green dash-dotted line in Figure~\ref{fig:fig4}. The apparent roll-off at high $k$ arises from the effective Nyquist limit imposed by the chosen pixel size, which ensures adequate sampling but slightly suppresses power near the resolution limit.

\subsection{Foregrounds removal} \label{sec:foregrounds}
Distinguishing the contributions of Galactic and extragalactic sources from the He~{\small{II}} or H~{\small{I}} signal poses a significant challenge in contemporary radio astronomy.
Foreground emissions typically exhibit a smooth frequency dependence, unlike the fluctuating 3.46~cm signal, which demonstrates poor correlation in frequency space. Additionally, the high rest-frequency of the helium hyperfine transition helps mitigate contamination from synchrotron radiation and the terrestrial ionosphere \citep[][]{Takeuchi+14}. These factors enable the development of pipelines specifically designed to extract the He~{\small{II}} signal from observations. We do not simulate the foregrounds contribution to the signal (only for then to remove it using such techniques) but we provisionally assume that the foreground cleaning technique removes only the mean of the signal (the $\mathbf{k} = 0$ mode of the Fourier transformed signal) while it leaves untouched the fluctuations around it.

While the dominant foregrounds in 1.42~GHz LIM are spectrally smooth (e.g., synchrotron and free-free emission), line-intensity mapping at 8.67~GHz may be affected by spectral confusion from line interlopers. In particular, low redshift HI emission can overlap in observed frequency with the He~{\small{II}} 3.46 cm line at $z > 5$, while molecular lines (e.g., OH) and hydrogen recombination lines may introduce additional fluctuations. These interlopers are not necessarily spectrally smooth and could therefore survive foreground-cleaning techniques based on low-rank subtraction or high-pass filtering. While some of these contaminants (e.g., low-z HI) may be identifiable due to their brightness and clustering properties, their removal is non-trivial and depends on both the instrument sensitivity and survey design.
In addition, man–made radio-frequency interference (RFI), including satellite broadcasting and communication signals in the UHF band ($\sim$0.3--3\,GHz), may further contaminate observations of the He\,\textsc{ii} line at $z\simeq 4$--10.

A comprehensive treatment of line interlopers (e.g., via cross-correlation techniques or blind line separation algorithms) is beyond the scope of this work but will be crucial for future detection efforts. For the purposes of this forecast, we conservatively assume HeII dominates the line signal, and defer detailed modeling of line confusion to future work.

\subsection{Single-dish thermal noise} \label{sec:thermal_noise_sd}
The observed signal is contaminated by the radio telescope thermal noise, which is Gaussian to first approximation. The rms noise fluctuation depends on the system temperature $T_{\rm sys}$, the integration time per pointing $t_{\rm pix}$ and the channel bandwidth, as
\begin{equation} \label{eqn:noise}
    \sigma_\mathrm{SD} (z) = \dfrac{T_{\rm sys}(z)}{\sqrt{t_{\rm pix} \Delta \nu}}\,.
\end{equation}

For SKA-1 MID, $T_{\rm sys}$ is given by the sum of several components and strongly depends on frequency \citep{SKAredbook}, as
\begin{equation}
    T_{\rm sys}(\nu) = T_{\rm spl} + T_{\rm CMB} + T_{\rm gal}(\nu) + T_{\rm rx}(\nu),
\end{equation}
where $T_{\rm spl} \approx 3\, \rm{K}$  and $T_{\rm CMB} \approx 2.73 \, \rm{K}$ denote the spill-over and the cosmic-microwave-background contributions, respectively.
The Galaxy contribution can be modelled as
\begin{equation}
    T_{\rm gal}(\nu) = 25 \left(\dfrac{408\, \rm{MHz}}{\nu} \right)^{2.75}\, \rm{K}
\end{equation}
and the receiver-noise temperature as
\begin{equation}
    T_{\rm rx}(\nu) = 15\,\rm{K} + 30 \left(\dfrac{\nu}{\rm{GHz}} - 0.75 \right)^{2}\, \rm{K}.
\end{equation}
This gives very large values of the system temperature for low redshifts (e.g. $T_{\rm sys} (z = 3) \approx 81$ K), a minimum around $z \approx 8.5$ ($T_{\rm sys} (z = 8.5) \approx 24$ K) and a gentle rise for higher redshifts (e.g. $T_{\rm sys} (z = 11) \approx 26$ K). It follows that, for fixed values of $t_\mathrm{pix}$ and $\Delta \nu$, the level of thermal noise is minimized for intermediate redshifts.
For DSA-2000 and PUMA, the system temperature is approximated to $25$ K \citep{Hallinan+19,Slosar+19} independently of frequency.

The integration time $t_\mathrm{obs}$ needed to map a portion of the sky over the solid angle $\Omega_\mathrm{surv}$ is given by
\begin{equation}
    t_\mathrm{obs} = \frac{t_\mathrm{pix}}{N_\mathrm{dish}\, N_\mathrm{beam}} \frac{\Omega_\mathrm{surv}}{\Omega_\mathrm{pix}}\;,
    \label{eqn:ttot}
\end{equation}
where $N_\mathrm{dish}$ is the number of antennas (see Table \ref{tab:radiosurveys}), each with $N_\mathrm{beam}$ (= 1 for the surveys we considered) feedhorns.

The analytical expression for the variance of the power spectrum due to the thermal noise, defined in equation \eqref{eqn:noise} is given by
\begin{align}
    N_\mathrm{SD} (z) & = \sigma_\mathrm{SD}^2 V_\mathrm{pix} = \nonumber \\ 
    & = \dfrac{T_\mathrm{sys}^2(z)}{\Delta\nu \,t_\mathrm{obs}\,N_\mathrm{dish}} A_\mathrm{surv}(z) \,l_\mathrm{pix},
    \label{eqn:noise_sd}
\end{align}
where $A_\mathrm{surv}$ is the area covered by the survey in physical units (while we denote as $\Omega_\mathrm{surv}$ the corresponding solid angle) and
$l_\mathrm{pix}$ is the physical length of the channel width (here chosen to be equal to $\Delta \chi$ as defined in Section \ref{sec:freq_band}).

\subsection{Interferometer thermal noise}\label{sec:thermal_noise_int}

In an interferometric setup, two antennas separated by a baseline of length $d$ measure a visibility $V(\mathbf{u}, \nu)$, where $u = |\mathbf{u}| = d / \lambda$ denotes the baseline projected onto the $uv$-plane. The resolution in $uv$-space is governed by the instrument's field of view (FOV), which, for an array of dishes, corresponds to the beam solid angle of an individual dish. This implies a typical resolution of $\delta u,\delta v \sim 1/\mathrm{FOV} \sim A_e / \lambda^2$, where $A_e$ is the effective area of a single dish ($A_e \simeq \pi(D/2)^2$). Visibilities that are separated by more than this scale in $uv$-space can be treated as statistically uncorrelated. For scales smaller than $A_e/\lambda^2$, one can average all visibilities within each resolution element of area $\delta u,\delta v$. This averaging reduces the noise by the number of visibilities per element, while the sky signal remains effectively unchanged (provided the $uv$ sampling is sufficiently fine). The noise in each $uv$ resolution element is then
\begin{equation}
    \sigma_\mathrm{int} (\mathbf{u},\nu) = \dfrac{T_\mathrm{sys}(z)}{\sqrt{t_\mathrm{obs}\Delta\nu}} \dfrac{\left[\lambda_\mathrm{\HeII} (1+z)\right]^2}{A_e} \dfrac{1}{\sqrt{N_b(\mathbf{u})}},
\end{equation}
where $N_b(\mathbf{u})$ ($=N_b(u)$ due to symmetric arguments) is the number density of baselines. If we assume that $N_b(\mathbf{u})$ is constant on the $uv$ plane between some $u_\mathrm{min} = D/(\lambda_\mathrm{\HeII} (1+z))$ and $u_\mathrm{max} = B/(\lambda_\mathrm{\HeII} (1+z))$, where $B$ is the core baseline of the survey (i.e. where the largest fraction of antenna is concentrated), we can write
\begin{equation}
\pi(u_\mathrm{max}^2-u_\mathrm{min}^2)N_b(u) = N_\mathrm{dish}(N_\mathrm{dish}-1)/2 \simeq N_\mathrm{dish}^2/2.
\end{equation}
Note that this is not the same as assuming an uniform distribution of antennas.

Following \citet{Bull+15} and references within \citep[e.g.][]{McQuinn06,Geil11,Villaescusa-Navarro14}, the power spectrum of the detector noise for interferometric observations, analog to the one defined in equation \eqref{eqn:noise_sd}, reads

\begin{equation}
    N_\mathrm{int}(z) \simeq 7 \left(\dfrac{B}{D} \right)^2 \dfrac{T^2_\mathrm{sys}(z)}{N^2_\mathrm{
    dish}} \dfrac{1}{\Delta\nu} \dfrac{A_\mathrm{FOV}}{t_\mathrm{obs}} \left[\dfrac{\Delta\nu}{\nu_\mathrm{HeII}} \dfrac{c(1+z)^2}{H(z)} \right], \label{eqn:noise_int}
\end{equation}
where $A_\mathrm{FOV} \sim \theta^2_\mathrm{FWHM}$ is the survey area (that corresponds to the field of view for an interferometer). The squared-brackets factor corresponds to the channel width in physical units.

\subsection{$k$-modes selection}\label{sec:kmodes}

It is important to carefully select the relevant scales when designing a radio survey, whether operated in single-dish mode or as an interferometer. In the former case, the accessible scales are limited by the angular resolution of each antenna ($k_\mathrm{FOV}$) and by the total area covered in each pointing ($k_\mathrm{area}$). In contrast, interferometers—thanks to their higher angular resolution—can access larger $k$-modes, constrained by the instrument’s field of view ($k_\mathrm{FOV}$) and by the maximum baseline length ($k_\mathrm{B}$). These characteristic scales depend on the specific instrument and the redshift under consideration, and are defined as\footnote{Assuming a flat universe, for which $(1+z)\,d_{\mathrm{a}}=\chi$.}:

\begin{align}
k_\mathrm{B} (z) & \simeq \dfrac{2\pi}{\chi(z)}\dfrac{B}{\lambda_\mathrm{HeII} (1+z)}, \label{eqn:kb} \\
k_\mathrm{FOV} (z) & \simeq \dfrac{2\pi}{\chi(z)}\dfrac{D_\mathrm{dish}}{\lambda_\mathrm{HeII} (1+z)}, \\
k_\mathrm{area} (z) & \simeq \dfrac{2\pi}{\chi(z)}\dfrac{1}{\sqrt{\Omega_\mathrm{area}}}. \label{eqn:ka}
\end{align}

For example for SKA-1 MID, we find $k_\mathrm{B} (z=4) = 7.29\,h\,\mathrm{Mpc}^{-1}$ and $k_\mathrm{FOV} (z=4) = 0.11\,h\,\mathrm{Mpc}^{-1}$. In this work, we assume an area per pointing\footnote{The antenna of a radio survey operating in single-dish mode observes only a limited sky region per pointing. The total survey area, $A_{\rm surv}$, can nevertheless be made arbitrarily large by combining multiple
pointings over the full observing time.}
of $\Omega_\mathrm{area} = 4\times 10^{-3} \,\mathrm{sr}\, (=13 \deg^2)$, corresponding to $k_\mathrm{area} (z=4) = 0.02\, h\,\mathrm{Mpc}^{-1}$. These scales are shown in Figure \ref{fig:fig4} as black vertical lines and denote the range of sensitivity for the two configurations. The power spectrum measured from the simulations considered in this work is limited by the size of the box ($k_\mathrm{F} = 0.0628 \, h\,\mathrm{Mpc}^{-1}$) and by its resolution, and it does not cover the all range of wavelengths needed to evaluate both configurations ($k_\mathrm{area} \leq k \leq k_\mathrm{FOV}$ for single-dish mode and $k_\mathrm{FOV} \leq k \leq k_\mathrm{B}$ for an interferometer). Moreover, while in principle $k_\mathrm{max} = 2\pi/(L/N_\mathrm{grid}) = 2\pi/(100\,h^{-1} \,\mathrm{Mpc}/128) = 8.04 \, h\,\mathrm{Mpc}^{-1}$, resolution effects are noticed already from $k\sim 4 \, h\,\mathrm{Mpc}^{-1}$ (for this reason the power spectrum is measured and presented in Figure \ref{fig:fig4} up until $k \simeq 3 \, h\,\mathrm{Mpc}^{-1}$). How to extend the power spectrum measured from simulations to the wavelengths of interests for this work is the subject of the next sections.

Equations~\ref{eqn:kb}-\ref{eqn:ka} specify the transverse wavenumber limits $k_\perp$ set by the baselines, dish size, and survey area. The radial modes, $k_\parallel$, are instead fixed by the channel width and total bandwidth. For the frequency resolutions considered here, the sampling in $k_\parallel$ is much denser than in $k_\perp$, so that $k_\parallel$ can be treated as effectively continuous. Consequently, the transverse limits $k_{\perp,\min}$ and $k_{\perp,\max}$ can be used as the effective bounds on the total wavenumber.

It is plausible that extending the integration over a broader range of scales, particularly at low-$k$ for single-dish and high-$k$ for interferometers, would marginally increase SNR, especially for instruments like DSA-2000 and PUMA, whose configurations offer access to both regimes. Future forecasts should incorporate full instrument sensitivity curves overlaid on the simulated power spectrum to optimize $k$-space coverage.

\subsubsection{Single-dish configuration}

As mentioned already in Section \ref{sec:powerspectrum}, the 3.46 cm signal is affected not only by the ionization state of helium but also by the underlying dark matter density. This last contribution leads for small $k$-modes, while the the \HeIII bubbles seems boosting the signal at higher $k$-modes. In the attempt of disentangle the two contribution, we compute the linear matter power spectrum with \textsc{CAMB} \citep{Lewis+00} and we compare it with the power spectrum of the 3.46~cm signal after a properly rescale (by a factor $C$, depending on the redshift considered). 
The matching of the two power spectra on large scales ($k \gtrsim 0.06 h \mathrm{Mpc}^{-1}$) is shown in Figure \ref{fig:fig4} as the orange continuous line. In order to gain access to scales larger than the one available with our simulations, we extend the brightness temperature power spectrum on the left-hand side, i.e. for scales $0.015\,h \mathrm{Mpc}^{-1} < k < 0.06 \,h \mathrm{Mpc}^{-1}$.
The orange dotted line in Figure \ref{fig:fig4} exhibits the convolution between the scaled theoretical power spectrum and the window function implementing the smoothing in the directions parallel and perpendicular to the line-of-sight, $P(k)_\mathrm{CAMB} \cdot W(k,\Delta\chi,\Sigma)$. The window function reads \citep{Li+16}
\begin{equation}
W(k)=e^{-k^2 \Sigma^2} \int_0^1 e^{-k^2\left(\Delta\chi^2-\Sigma^2\right) \mu^2} \mathrm{d} \mu\, ,
\end{equation}
where $\mu = \cos \theta$ with $\theta$ the spherical polar angle in $k$-space.
We display in Figure \ref{fig:fig4} as red points the selected $k$-values extrapolated from the smoothed rescaled matter power spectrum.

\subsubsection{Interferometer configuration}

The largest $k$-mode accessible from simulations is set by their resolution, but it is limited to $k_\mathrm{cut-off}\simeq 3 \, h\,\mathrm{Mpc}^{-1}$ (corresponding to a scale of $\sim 2\,h^{-1}$ Mpc) to avoid resolution effects in the computation of the power spectrum. We notice, from the analysis of the power spectrum, that these scales are well traced by the matter power spectrum when accounting for a small boost due to the \HeIII bubbles. We extrapolate the $k$-modes between $k_\mathrm{cut-off}$ and $k_\mathrm{B}$ by fitting a power-law to the power spectrum at large $k$-modes, and sampling using the same $\delta k$. These $k$-values are shown as dark blue points in Figure \ref{fig:fig4}.

\subsection{Clustering and shot-noise regimes of the He II power spectrum} \label{sec:clustering}

The physical interpretation of the 3.46\,cm power spectrum depends on the spatial scales probed. On large scales (low $k$), fluctuations predominantly trace the clustering of He\,\textsc{ii} regions around AGN and broadly follow the underlying matter density field. On small scales (high $k$), the signal becomes dominated by the discreteness of these ionized structures and by the strong concentration of emission within overdense gas, in qualitative analogy with the shot-noise regime familiar from mm-wave and 21~cm line-intensity mapping. This scale dependence has direct implications for observations: interferometric arrays are intrinsically more sensitive to the small-scale, shot-noise-dominated component of the signal, while single-dish surveys preferentially probe the large-scale clustering power.

The transition scale depends on the spatial distribution and duty cycle of AGN. While this behavior is well known in mm-wave and 21~cm LIM \citep[e.g., ][]{Breysse17,Uzgil19}, it has not been explicitly studied for the HeII hyperfine transition. Our simulations suggest that the HeII signal in overdense regions leads to enhanced power at small scales, consistent with a hybrid clustering–shot-noise regime.

\begin{figure*}
    \centering
\includegraphics[width=0.9\textwidth]{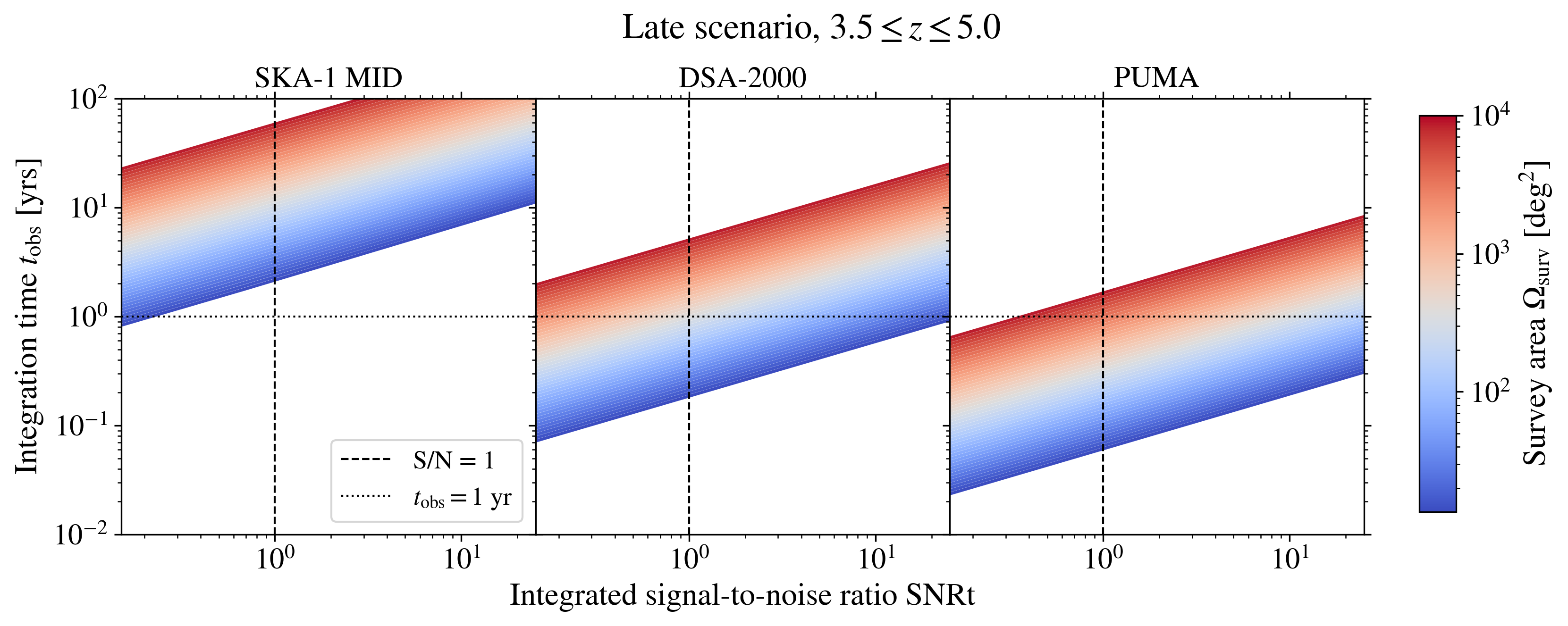}
    \caption{Relation between the integration time $t_\mathrm{obs}$ and the integrated signal-to-noise ratio SNRt for the redshift range $3.5 \leq z \leq 5$ for single-dish observations. Different colors represent different survey areas $\Omega_\mathrm{surv}$, while the dotted black line shows a reference integration time of 1 year, and the dashed back lack line shows a reference SNRt of 1. Each panel corresponds to a unique survey.}
    \label{fig:fig5}
\end{figure*}

\begin{figure*}
\centering
\includegraphics[width=0.9\textwidth]{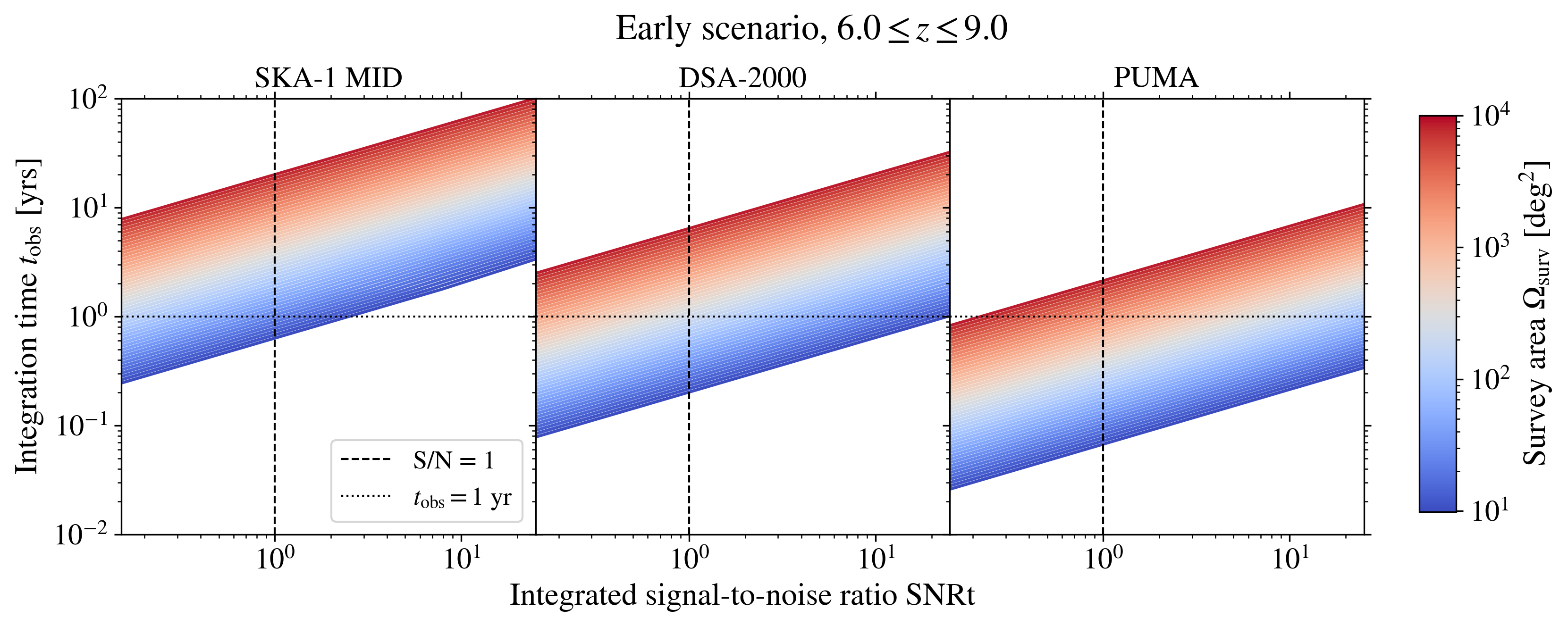}
\caption{As Figure \ref{fig:fig5}, for the redshift range $6 \leq z \leq 9$ assuming an early reionization scenario.} \label{fig:fig6}
\end{figure*}

\begin{figure}
    \centering    \includegraphics[width=\columnwidth]{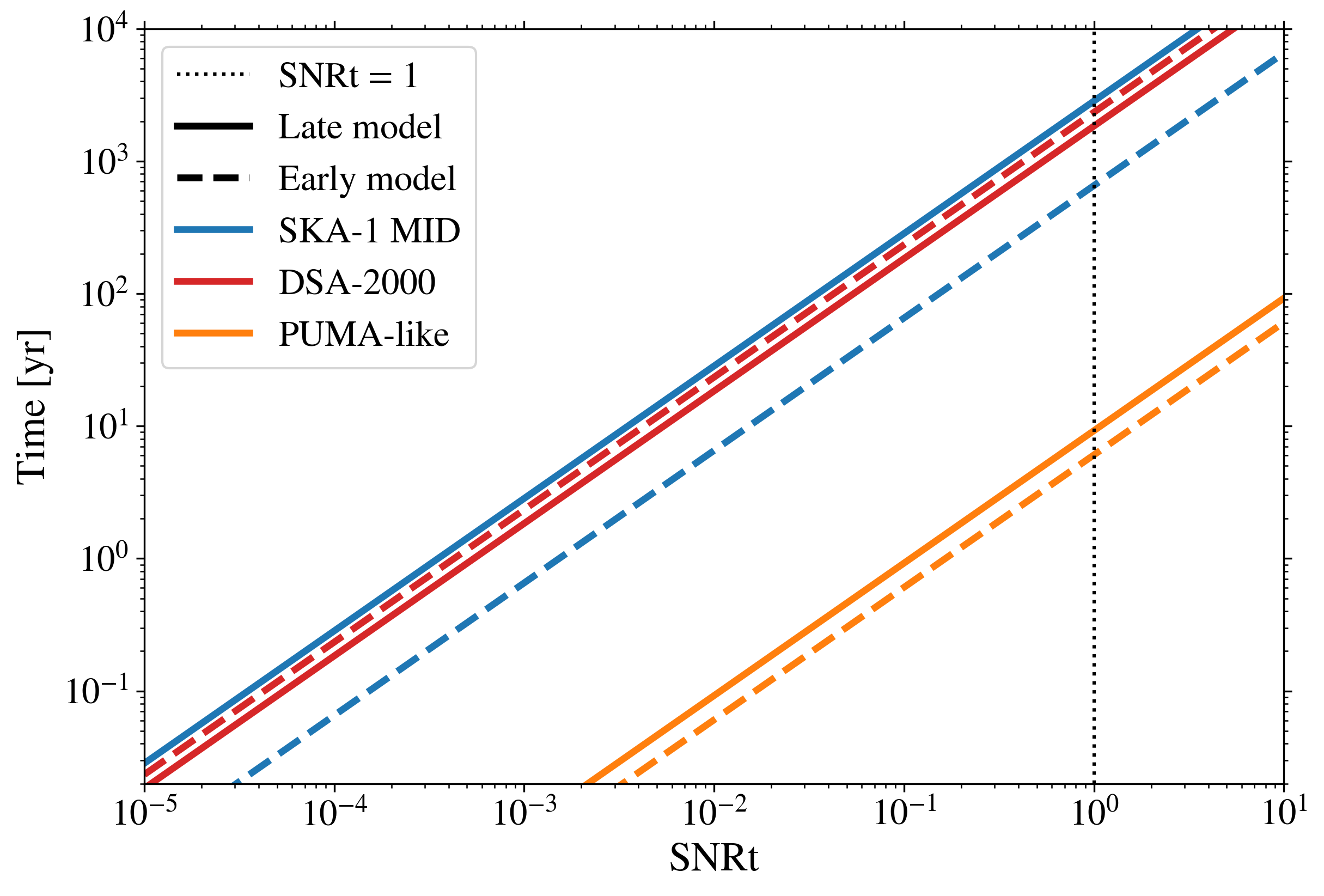}
    \caption{Relation between the integration time and the integrated signal-to-noise ratio SNRt for interferometric observations. Forecasts are shown for the three radio surveys considered, SKA-1 MID (blue), DSA-2000 (red), PUMA-like (orange), and for the two reionization model studied in this work, Late (continuous lines) and Early (dashed lines) scenarios. The black dotted line marks a SNRt of 1.}
    \label{fig:interf}
\end{figure}

\section{Signal-to-noise ratio} \label{sec:signal-to-noise}

In our pursuit of assessing the detectability of the signal within current radio astronomy surveys, it is helpful to define the signal-to-noise ratio for the power spectrum as
\begin{equation}
     \mathrm{SNR}^2(k) = \dfrac{P^2(k)}{\sigma_P^2(k)}  = N_q \dfrac{P^2(k)}{\left[P(k) + N \right]^2},
     \label{eqn:SNR}
\end{equation}
where $N$ denotes the power spectrum of the detector noise for either single-dish ($N_\mathrm{SD}$, defined in equation \eqref{eqn:noise_sd}) or interferometric configurations ($N_\mathrm{int}$, defined in equation \eqref{eqn:noise_int}), and the number of modes within the $k$-bin of width $\delta k$ is 
\begin{equation}
    N_q = \dfrac{1}{2} 4\pi k^2 \delta k \dfrac{V_\mathrm{surv}}{(2\pi)^3} = \dfrac{k^2 \delta k}{(2\pi)^2} V_\mathrm{surv}\, ,
    \label{eqn:Nq}
\end{equation}
where $V_\mathrm{surv} = A_\mathrm{surv} l_\mathrm{s}$ is the volume of the survey given by the observed area perpendicular to the line-of-sight $A_\mathrm{surv}$ and the comoving distance in the direction parallel to the line-of-sight $l_\mathrm{s}$. 

The standard deviation on the power spectrum, as defined in equation \eqref{eqn:SNR}, accounts for the contribution of the number of independent modes within the $i$-th bin (the higher it is, the smaller is the error on the power spectrum) and of the white noise introduced by the instrument. 

We are interested in quantifying the required integration time to detect the He~{\small{II}} signal, hence we express the SNR as a function on $t_\mathrm{obs}$ by combining equations~\eqref{eqn:ttot}, \eqref{eqn:noise_sd}, \eqref{eqn:SNR}, \eqref{eqn:Nq} for a single-dish configuration
\begin{equation}
    \mathrm{SNR}_\mathrm{SD}^2 (k) = \dfrac{k^2 \delta k}{(2\pi)^2} A_\mathrm{surv} l_\mathrm{s} \left[1 + \dfrac{1}{P(k)} \dfrac{T_\mathrm{sys}^2 A_\mathrm{surv} l_\mathrm{pix}}{t_\mathrm{obs}N_\mathrm{dish} N_\mathrm{beam} \Delta\nu} \right]^{-2}.
    \label{eqn:stn_sd}
\end{equation}
Note that for a given instrument, at a given redshift, at a given scale, equation \eqref{eqn:stn_sd} reads 
$\mathrm{SNR} \propto l_\mathrm{s}^{1/2} A_\mathrm{surv}^{-1/2} t_\mathrm{obs}$.
Therefore, an optimal SNR is achieved when observing smaller areas but wider redshift ranges.
For an interferometer, equation \eqref{eqn:stn_sd} reads
\begin{multline}
\mathrm{SNR}_\mathrm{int}^2 (k) = \dfrac{k^2 \delta k}{(2\pi)^2} A_\mathrm{surv} l_\mathrm{s} \\ \times \left[1 + \dfrac{1}{P(k)} 7\left(\dfrac{B}{D}\right)^2\dfrac{T_\mathrm{sys}^2 A_\mathrm{FOV} \lambda_\mathrm{HeII}}{t_\mathrm{obs}N_\mathrm{dish} N_\mathrm{beam}} \dfrac{(1+z)^2}{H(z)} \right]^{-2}.
\label{eqn:stn_int}
\end{multline}

To improve detection capabilities, it is advantageous to consider the integrated signal-to-noise ratio, denoted as SNRt, obtained by combining the contributions from all scales, defined as
\begin{align}
    \mathrm{SNRt}^2 & = \sum_k \left(\dfrac{P(k)}{\sigma_{P(k)}}\right)^2.
    \label{eqn:SNRt}
\end{align}

Since the power spectrum varies with redshift, we utilize the mean power spectrum within the specified redshift range ($3.5\leq z \leq 5$ for the Late model, $6\leq z \leq 9$ for the Early model) for our following analysis. We have presented in Section \ref{sec:kmodes} and in Figure \ref{fig:fig4} the $k$ values used in the calculation of SNRt (it is worth noting that the same procedure is applied to other redshifts). These values include both those derived from the simulation boxes and those obtained from the scaled matter power spectrum.

The relationship between the integration time $t_\mathrm{obs}$ and the integrated signal-to-noise ratio, SNRt, is presented in Figures~\ref{fig:fig5} and \ref{fig:fig6} for single-dish observations, and in Figure~\ref{fig:interf} for interferometric observations.

For single-dish setups, we show results for a fixed line-of-sight depth, defined by the redshift ranges $3.5 \leq z \leq 5$ (Late model) and $6 \leq z \leq 9$ (Early model), and various survey areas $\Omega_\mathrm{surv}$ across the three instruments considered. Optimal performance is obtained when observing a relatively compact region of the sky over an extended frequency range. Larger survey areas require significantly longer integration times to reach the same $\mathrm{SNRt}$, since the time per pixel decreases when spreading the observing time over a wider area. Among the three surveys, a PUMA-like configuration, thanks to its large number of antennas and intermediate dish size, provides the highest sensitivity. While SKA-1 MID and DSA-2000 would require prohibitively long observations to reach $\mathrm{SNRt} \sim 1$, a PUMA-like survey in single-dish mode could achieve a marginal detection of the 3.46\,cm signal ($\mathrm{SNRt} \sim \mathrm{few}$) within 
$\lesssim 1000\,\mathrm{h}$, in both reionization scenarios. Next-generation instruments may be able to access the 
He\,\textsc{ii} hyperfine signal for the first time.

For interferometric observations all considered instruments require prohibitively long integration times, though the PUMA-like survey again performs best for both the Early and Late models. Compared to single-dish mode, interferometry is more sensitive to small-scale fluctuations where the He~{\small II} signal peaks, but it also requires longer observations to reach a comparable SNRt.

The underlying difficulty lies in the intrinsic faintness of the 3.46\,cm signal. Because the spin temperature in low-density regions is only weakly coupled to the kinetic temperature, the differential brightness temperature is suppressed across much of the intergalactic medium. As a result, the signal is mostly generated in overdense regions, which coincide with the scales where instrumental noise and resolution limits are most restrictive. 

Despite the low SNRt, our simulations show that the Early and Late reionization models produce distinguishable power spectra. In the Early model, the presence of high-redshift quasars leads to an elevated He~{\small II} fraction and a stronger signal at $z \gtrsim 5$, with a relatively flat power spectrum at small scales. In the Late model, the signal builds up more slowly and peaks at $z \sim 4$, with a steeper shape and greater power at intermediate scales. These differences, though apparent in simulation, are effectively unobservable with current instrument sensitivities.

\section{Summary \& conclusion}  \label{sec:results}

In this work, we assessed the detectability of the 3.46\,cm hyperfine transition of singly-ionized helium with upcoming and proposed radio surveys, using hydrodynamical simulations post-processed with radiative transfer to model Early and Late helium reionization scenarios. We computed the corresponding He\,\textsc{ii} power spectra and evaluated the performance of SKA-1 MID, DSA-2000, and a PUMA-like array in both single–dish and interferometric observing modes.

Our results show that the intrinsic faintness of the 3.46\,cm signal (driven also by weak spin–temperature coupling in most of the IGM) suppresses the brightness temperature and limits detectability for current surveys. Although the Early and Late models produce distinct power spectra in simulation, the predicted amplitudes are well below the thermal-noise levels of SKA-1 MID and DSA-2000, both in single-dish and interferometric setups. 

A PUMA-like survey in a single-dish setup, however, could detect the 3.46\,cm signal with a SNRt of a few in $\lesssim 1000\,\mathrm{h}$. Such a detection would constitute a first step toward accessing the 3.46\,cm signal, and even distinguish between Early and Late helium-reionization scenarios, constraining the sources of helium reionization. Future progress will likely require improved sensitivity, denser baseline coverage, larger bandwidths, or complementary approaches such as global-signal measurements or cross-correlations with galaxy surveys. Likewise, alternative statistics sensitive to higher-order moments and to the global geometry and topology of the He\,\textsc{ii} field may further help differentiate reionization models. Larger-volume simulations would also be beneficial to better capture the accessible $k$-range and reduce sample variance on large scales.

\begin{acknowledgements}
BS carried out part of this work while being part of the International Max Planck Research School in Astronomy and Astrophysics and the Bonn Cologne Graduate School. BS and CP gratefully acknowledge the Collaborative Research Center 1601 (SFB 1601 sub-project C6) funded by the Deutsche Forschungsgemeinschaft (DFG, German Research Foundation) – 500700252. BS and SEIB
are supported by the Deutsche Forschungsgemeinschaft (DFG) under Emmy
Noether grant number BO 5771/1-1.
CP is grateful to SISSA, the University of Trieste, and IFPU, where part of this work was carried out, for hospitality and support.
We thank Fabian Walter, Anna Pugno and Elena Marcuzzo for the useful comments and discussions. 
\end{acknowledgements}

\bibliographystyle{aa}
\bibliography{aa}

\clearpage

\end{document}